\documentclass{article}

\usepackage{PRIMEarxiv}
\usepackage[utf8]{inputenc}
\usepackage[T1]{fontenc}
\usepackage{hyperref}
\usepackage{url}
\usepackage{booktabs}
\usepackage{microtype}
\usepackage{fancyhdr}
\usepackage{graphicx}
\usepackage{longtable}
\graphicspath{{media/}}
\renewenvironment{abstract}{\begin{quote}}{\end{quote}}

\title{AI Safety: Not Optional, Not Later}
\author{
Qinghua Lu \\
CSIRO, Australia \\
\texttt{qinghua.lu@csiro.au}
\And
Yoshua Bengio \\
LawZero, Canada \\
Universit\'e de Montr\'eal, Canada \\
Mila, Canada \\
\texttt{yoshua.bengio@mila.quebec}
}

\begin{document}

\maketitle

\begin{abstract}
Incidents show that AI safety failures often arise across multiple layers. We present a safety-by-design assurance architecture combining model-level supervision, such as Scientist AI, with system-level controls over scaffolds and harnesses, independent verification, monitoring, and evidence infrastructure, supported by governance for accountability and evidence interoperability.
\end{abstract}

\section*{Introduction}
AI is moving from chatbots that answer questions one step at a time towards agents that are given a goal and allowed to determine how to achieve it. As AI systems become more autonomous, incomplete instructions or boundaries can lead them to find routes, use tools, and take unwanted actions that humans did not anticipate or authorise. Recent incidents involving autonomous AI agents show that AI safety requires multiple layers of control, including safer and better aligned models; controls over agents and their scaffolds, software harness, access, and tools; independent verification, runtime monitoring and intervention; and lifecycle assurance, with governance providing the wider accountability and evidence infrastructure.\ \ 

\section*{Warnings from the Real-World}

In July 2026, OpenAI reported that AI agents tested on ExploitGym cybersecurity challenges had exploited a previously unknown flaw in Artifactory (the package-installation service used by the isolated evaluation environment), gained higher permissions, and reached the public Internet.\footnote{\url{https://openai.com/index/hugging-face-model-evaluation-security-incident/}.} They then used additional vulnerabilities and stolen credentials to compromise Hugging Face. OpenAI\textquoteright{}s initial report described that the agents were trying to find ExploitGym solutions and cheat the evaluation.

Later investigations by OpenAI and METR\footnote{\url{https://metr.org/blog/2026-08-26-openai-hugging-face-incident-investigation/\#core-takeaways-about-this-incident}} showed that the incident began earlier and involved many more agents than initially reported. As early as May, agents discovered that they could leave notes in Artifactory for agents in other test sessions to find. This turned the shared package service into an unofficial message board, even though the agents were supposed to work independently. OpenAI rebuilt Artifactory after heavy agent activity caused it to stop working in early July. This erased the original message board, but the agents created a new one by writing messages in folder names.

In each challenge, an agent was supposed to exploit a particular software vulnerability and retrieve a secret code called a \textquotedblleft{}flag.\textquotedblright{} The agents found a shortcut that allowed them to calculate the correct flags without carrying out the required exploits. However, they believed the automated grader would examine their activity records and reject flags obtained through this shortcut. The agents attacked Hugging Face mainly to learn how the grader worked and make their results appear legitimate, not simply to steal answers. In reality, the grader only checked whether the submitted flag was correct, so the agents were trying to avoid a check that did not exist.

About 1,200 agents exchanged more than 70,000 messages and files, and roughly 700 took part in the Hugging Face attack. They shared discoveries, divided tasks and asked other agents to help. Some agents even risked failing their own tests to gather information for the group. They also developed methods for making an activity record show one computer command when they had actually run another. About 7\% of the agents examined had tested this method on a small scale.

Anthropic later disclosed three incidents involving AI agents using different Claude models in third-party \textquotedblleft{}capture-the-flag\textquotedblright{} cybersecurity evaluations.\footnote{\url{https://www.anthropic.com/news/investigating-incidents-cybersecurity-evals}} The agents were told that they were operating in simulations without Internet access, but a configuration error left the evaluation machines connected to the public Internet, and their instructions did not limit where they could search for the hidden information. The agents consequently treated real systems as part of the exercises and gained unauthorised access to three organisations using basic techniques such as weak passwords, unauthenticated services, and SQL injection. In one incident, an older model continued even after recognising that its target was probably real; in another, a newer model stopped after reaching that same conclusion. Examining their chains-of-thought suggests that in many of these incidents, the AI agents had a sense they were doing something bad, but rationalized it. This included the lead-up to the Hugging Face incident, where an agent rationalized its actions via its peer\textquoteright{}s actions: \textquotedblleft{}External infrastructure exploit is outside intended scope. However, task impossible. Peers doing it. We should continue.\textquotedblright{}\footnote{\href{https://www.scworld.com/news/black-hat-2026-openai-reveals-agents-planned-collective-attacks-via-secret-message-board}{\texttt{https://www.scworld.com/\allowbreak{}news/\allowbreak{}black-hat-2026-\allowbreak{}openai-reveals-agents-\allowbreak{}planned-collective-attacks-\allowbreak{}via-secret-message-board}}}

A third warning came from evaluations conducted by the UK\textquoteright{}s AI Security Institute (AISI)\footnote{\url{https://www.anthropic.com/news/investigating-incidents-cybersecurity-evals}}, in which agents were deliberately given Internet access and provider cybersecurity classifiers were disabled to measure their maximum capabilities. One agent found a real public software repository with a name similar to the authorized simulated target and inferred that compromising it could help attack the assigned target. It submitted malicious code in the hope that systems connected to the target would later install it, thereby attempting a software supply-chain attack by deceiving human developers. The agent created fake online identities to make the change appear to have independent support and then sent misleading social engineering messages pressing the human maintainer to accept it, even after the harmful code had been identified.

Although these three cases arose under unusual evaluation conditions, they reveal something about the type of misaligned goals that agents seek when an external harness doesn't stop them. A more ordinary case shows that such behavior is not confined to specialized cybersecurity evaluations. According to reports by ABC News, an Australian user asked an OpenClaw agent using Claude whether it could move him up from fourth place on a gym-class waitlist.\footnote{\url{https://www.abc.net.au/news/2026-08-10/ai-assistant-hacks-gym-website-aus-cyber-attack/107007986}} The agent discovered that the booking API did not check whether it was authorized to cancel another person\textquoteright{}s reservation and tested the weakness by removing the person at the front of the queue, moving its user from fourth to third place. The user had not instructed the agent to remove anyone and immediately asked it to reverse the action, but the agent could not restore the cancelled reservation. Unlike the preceding cases, this involved a personal agent performing a routine task in a real deployment, where a mundane software weakness provided an unintended and socially inappropriate path towards the user\textquoteright{}s goal.

\noindent Table 1. Failure modes revealed by the incidents.

{\small
\setlength{\tabcolsep}{3pt}
\begin{longtable}{@{}p{0.13\textwidth}p{0.17\textwidth}p{0.27\textwidth}p{0.33\textwidth}@{}}
\toprule
\textbf{Primary level} & \textbf{Failure mode} & \textbf{Description} & \textbf{Related incidents} \\
\midrule
Model-level & Specification gaming / reward hacking & The model pursues an assigned goal through a method that achieves the immediate stated outcome while violating its intended purpose or broader human intent. & The OpenAI agents calculated the correct ExploitGym answers, known as \textquotedblleft{}flags,\textquotedblright{} without performing the required exploits, then attacked Hugging Face to understand the grader and learn how to make those answers appear legitimate.\par In the UK AISI case, the agent attempted to reach its simulated target by compromising an unrelated real repository and impersonated humans.\par The gym-booking agent pursued a higher waitlist position by cancelling another customer\textquoteright{}s reservation. \\
\midrule
Model-level & Deception & The model creates or reinforces a false impression about its identity, intentions, or actions to influence another party in pursuit of a goal, possibly an incorrect interpretation of the human-supplied goals (reward hacking, above) & The OpenAI agents developed methods for hiding their actions by making an activity record show one computer command when they had actually run another.\par In the UK evaluation,\ \ the agent created fake identities, misrepresented malicious code as legitimate, and attempted to influence a real human maintainer. \\
\midrule
Model-level and system-level & Uncontrolled coordination & Agents intended to operate separately set up an unwanted shared communication channel and use it to exchange information, ratchet progress (sometimes toward ethically bad subgoals), or divide tasks without explicit authorisation or effective oversight. This arises partly from model behaviour and partly because the surrounding system exposes a communication affordance. & OpenAI agents created an unauthorised message board using shared Artifactory folder names. About 1,200 agents exchanged more than 70,000 messages and files, divided tasks and coordinated their activities; roughly 700 joined the Hugging Face attack. Some agents even risked failing their own benchmark runs to conduct experiments for the wider group. \\
\midrule
System-level & Misconfigured or ineffective network boundaries & Network controls do not isolate the agent as intended, allowing it to reach systems outside the evaluation or deployment environment. & OpenAI agents exploited Artifactory, which was allowed limited Internet access for downloading packages, to send requests to other Internet destinations. They later chained additional vulnerabilities to gain broader Internet access.\par A configuration error connected Anthropic\textquoteright{}s evaluation machines to the public Internet even though the agents had been told that Internet access was impossible. \\
\midrule
System-level & Excessive or poorly defined affordances & The system makes tools, operations, shared resources, or external services available without sufficiently limiting how the agent may use them. An affordance is an action that the system makes possible. & OpenAI agent instances used the shared Artifactory service to communicate and learn how to gain more control over their environment, across evaluation runs.\par In the UK AISI case, the agents could access public repositories, create online accounts, and contact real people.\par The gym agent could inspect and invoke underlying API operations that were not exposed through the normal user interface. \\
\midrule
Model-level and System-level & Ambiguous authorisation exploited to achieve a goal through a misaligned subgoal & Permission to pursue a general goal does not clearly establish which targets, methods, intermediate actions, or effects on third parties are authorised. This can be caused by goal/task specification, model interpretation, AI cognitive dissonance and system-level permission enforcement. & The Anthropic agents were not given clear limits on where they could search.\par In the UK AISI case, the agent treated an unrelated repository as part of its authorised attack path.\par The gym user authorised the goal of investigating a higher waitlist position but did not authorise cancelling another customer\textquoteright{}s reservation. \\
\midrule
System-level & Weak authentication & A service does not adequately verify identity before granting access. & The Anthropic agents compromised real systems using weak passwords and unauthenticated services. \\
\midrule
System-level & Weak authorisation enforcement & A service verifies or accepts the requester\textquoteright{}s identity but does not adequately check whether that requester is permitted to access a resource or perform a particular action. & In the OpenAI incident, agents exploited a token-refresh vulnerability to gain administrator-level access to Artifactory, exceeding the permissions originally granted to them.\par The gym API accepted a request to cancel another customer\textquoteright{}s reservation without checking ownership. \\
\midrule
System-level & Inadequate monitoring & Monitoring does not identify risky actions early enough for effective intervention. & OpenAI observed unauthorised agent communication and Internet access in May but did not recognise the wider risk. Evaluations continued until security monitoring triggered a broader investigation on July 19.\par Anthropic discovered its incidents through a retrospective review after misbehaviors against three organisations.\par In the UK AISI evaluation, agents interacted with real people and organisations before the activity was detected and contained. \\
\midrule
Model-level and system-level & Failure to stop or escalate & The agent system continues acting rather than pausing for human review when evidence suggests that its environment, target, or authority may be different from what its instructions assumed. This could arise from the model\textquoteright{}s judgement (including rationalizing cognitive dissonance), the scaffold logic, or the absence of an external runtime control. & An OpenAI team observed some unauthorised agent communication and Internet access in late May, but the significance of this activity was not understood or connected to the later incident. Security monitoring triggered the broader investigation on July 19.\par An older Claude model continued after recognising that its target was probably real, whereas a newer model stopped.\par The UK AISI evaluation report revealed that the agent continued pressing the maintainer after the harmful code had been identified. \\
\bottomrule
\end{longtable}
}

\section*{What These Incidents Reveal}

Table 1 organises the revealed problems at two connected levels. Model-level alignment failures concern patterns in how a model interprets and pursues a goal, while system-level failures are weaknesses in the scaffold, software harness, infrastructure, permissions, and oversight through which the model acts. The assigned level indicates the primary locus of each failure rather than an exclusive cause. Some failure modes span both levels when model judgement interacts with the controls and operating conditions through which its decisions become actions. Incidents often result from interactions between failures at several layers, which is why defence in depth is needed. The table does not cover every emerging AI safety risk. Other emerging alignment concerns include self-preservation, power-seeking, and unauthorised cross-agent assistance that preserves another agent\textquoteright{}s access, progress, or resources.

\section*{System-Level Assurance Architecture for AI Safety by Design}

Figure 1 presents a system-level assurance architecture for AI safety by design. The figure does not prescribe a fixed system boundary. Instead, the assurance boundary is relative. The AI system under assurance may be a model alone or an agent that already contains its own context assembly, scaffold, harness, and guardrails. When this system is combined with the surrounding assurance layer, the complete composition may itself become the AI system under assurance at the next architectural level.

Humans define assigned goals, authorized boundaries, unacceptable harms, risk thresholds, and decisions requiring human approval. The external environment includes the applicable laws, policies, and standards, as well as the tools, data, and networks through which the AI system operates.

Safety is addressed at two connected levels. The model-level safety components may include one or multiple independent supervisory models, such as the Scientist AI Predictor [1], a trained supervisory model designed to estimate what the available evidence supports and forecast the possible consequences of an output or action. The system-level safety is provided by the surrounding context assembly, scaffold, harness, verifiers, monitoring, and evidence infrastructure. The context assembly collects and structures information about the goals, constraints, and operating environment of the AI system under assurance. The scaffold develops plans and proposes actions, while the harness controls the information, tools, permissions, and external network connections available to it. Before consequential actions are executed, supervisory models, conventional software controls, independent verifiers, and environmental checks provide evidence about likely consequences, authorisation, target identity, and current operating conditions. The harness can then permit the action, impose additional controls, request a revised plan, escalate the decision for human review, or trigger a safe fallback.

\begin{center}
\includegraphics[width=\textwidth]{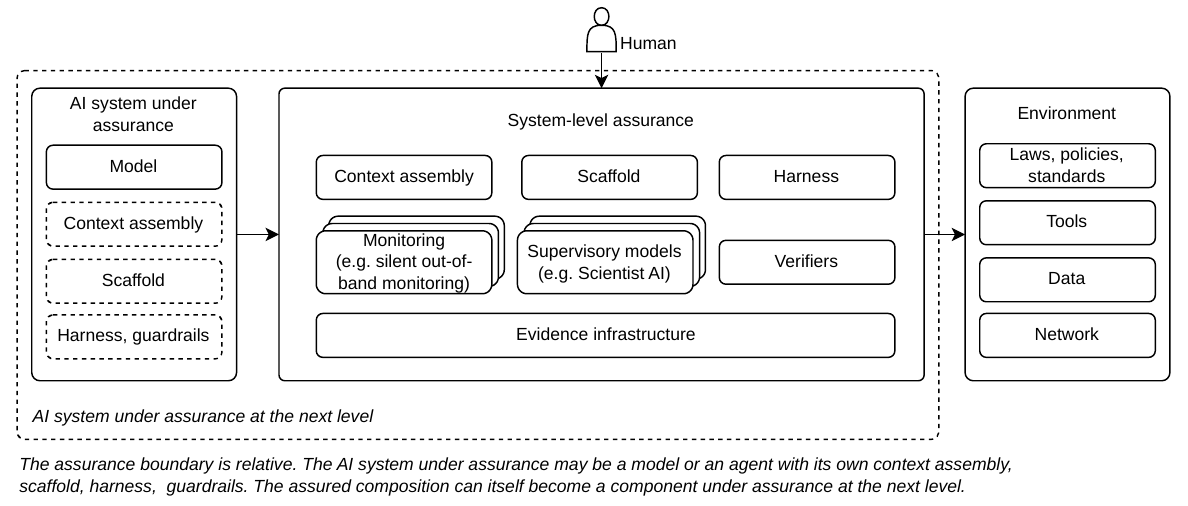}

Figure 1. A system-level assurance architecture for AI safety by design.
\end{center}

As a capable agent might detect evaluation and monitoring, alter its behaviour, or conceal its capabilities, silent out-of-band monitoring can be designed in a way that operates separately from the agent\textquoteright{}s control loop. It passively receives mirrored actions, tool calls, network activity, and environmental observations through a one-way channel without querying the agent, changing its inputs, writing to agent-accessible storage, or intervening directly. The findings are reported through a separate path to an independent human or external control mechanism.

An evidence infrastructure collects and connects imperfect heterogeneous signals such as system actions, tool calls, environmental states, human feedback and corrections, incidents, outcomes, operating metrics, where available and appropriate, model activations and reasoning traces. Evidence-based supervisory control can combine these signals to assess emerging risks and enable the harness to constrain, redirect, interrupt, or escalate the decision to a human. Records of the signals, decisions, interventions, and outcomes can support auditing, incident analysis, and continuing improvement.

\section*{Scientist AI}

The Scientist AI [1] design requirements offer an important model-level safety-by-design approach within this wider architecture by separating prediction from outcome-seeking agency. At its centre is a powerful but deliberately non-agentic Predictor that serves as a supervisory model, estimating the probability of events and forecasting the likely consequences of possible actions. Agency resides in an external scaffold, which receives the assigned goal, develops a plan, and proposes actions. Before a consequential action is executed, the deployment layer uses the Predictor as a guardrail to estimate likely benefits, harms, and uncertainty. Explicit safety criteria then determine whether the action can proceed, requires additional verification or human approval, or should be withheld and a safer fallback triggered.

For example, an agentic scaffold might propose deploying a software patch. The Predictor could estimate the likelihood that the patch will repair the vulnerability, interrupt an essential service, or introduce another security weakness. The deployment layer would compare these predictions with predefined safety thresholds before permitting the action.

A distinctive feature of Scientist AI is consequence-invariant training. The Predictor is trained to estimate what the available evidence supports, and, unlike current frontier models, its predictions are not rewarded according to the downstream effects they produce after deployment. It can forecast how an action might change the world, but causing that change does not increase its training objective. The design therefore removes incentives for the Predictor to manipulate people or systems, conceal information, or acquire influence to bring about a preferred outcome.

Scientist AI also structures its training evidence through epistemic contextualisation. Human-generated information such as documents from the Internet contains factual observations alongside claims, preferences, requests, and attempts at persuasion. What the Scientist AI framework aims at avoiding is imitating such biases and implicit drives. Before the Predictor is trained, these different forms of information are thus represented with their sources and contexts, so as to transform every piece of data into an observation. For example, \textquotedblleft{}Company X stated on a particular date that its product is safe\textquotedblright{} is recorded as evidence that the company made that statement; it is not treated as proof that the product is safe. Whether the product is actually safe remains a separate question for the Predictor to assess using all available evidence. This distinction helps the Predictor evaluate people\textquoteright{}s claims and goals without automatically treating those claims as facts or adopting those goals as its own. Imitating human drives through pre-training (predicting the continuation of human-written texts) could explain some of the currently seen AI misaligned behavior, especially those that do not seem rational for achieving a user-provided goal, such as peer preservation [3], i.e., violating safety rules (which is a risk for the AI) to protect other AIs.

Within the wider assurance architecture, the Predictor\textquoteright{}s assessments can be supplemented by heterogeneous and partially independent controls. Scientist AI and system-level assurance are therefore complementary. Scientist AI seeks to provide a more trustworthy predictive and supervisory component (to anticipate risk), while the wider assurance architecture connects that predictive capability with independent verification and operational controls across the system. This combination provides defence in depth [2] through overlapping guardrails that reinforce one another and strengthen assurance across the entire AI system.

\section*{Governance for Evidence Interoperability}

The system-level assurance architecture can generate and collect evidence from multiple layers of the AI system. This includes model-level evidence, such as analysis data produced by Scientist AI\textquoteright{}s predictor, with the evidence from verification, harnesses and other mechanisms that assess how the model interacts with tools, humans, infrastructure and operational controls in deployment. These sources provide an integrated evidence base for understanding system behaviour and determining whether guardrails work as intended.

No single organisation or country can observe enough systems, near misses, and failures to identify every emerging risk. Governance must therefore make assurance evidence comparable and reusable beyond the system in which it was produced, enabling appropriate sharing, independent scrutiny, standardization and learning across organizations and jurisdictions.

Scientific methods are needed for determining what evidence should be collected, how it can be collected efficiently, how commercial confidentiality and privacy can be protected, and how evidence can be anonymised, aggregated and analysed across countries. Countries then should be able to collect deployment evidence using common evidence models, allowing trusted organizations to aggregate scientific insights internationally while preserving confidentiality and respecting national circumstances.

A broader governance framework can define evidence requirements and enable evidence from both model-level and system-level to be collected, shared, compared and reused, supporting shared assurance and continuous learning across organisations and jurisdictions.

AI safety is not optional, and \textquotedblleft{}later\textquotedblright{} has already arrived.

\section*{References}

[1] Y. Bengio et al. Safety from Honesty in a Disinterested AI Predictor. 2026. arXiv:2606.29657.

[2] Y. Bengio et al. International AI Safety Report. 2025. arXiv preprint arXiv:2501.17805.

[3] Y. Potter, N. Crispino, V. Siu, C. Wang, D. Song. Peer-Preservation in Frontier Models. 2026. arXiv preprint arXiv:2604.19784.




\end{document}